\documentclass[12pt, draftclsnofoot, onecolumn]{IEEEtran}
\IEEEoverridecommandlockouts
\usepackage{cite}
\usepackage{amsmath,amssymb,amsfonts}
\usepackage{dblfloatfix}    
\usepackage{graphicx}
\usepackage{textcomp}
\usepackage{xcolor}
\usepackage{comment}
\usepackage{algorithm}
\usepackage{algpseudocode}
\usepackage{float}
\usepackage{nicefrac}       
\usepackage{bbm}
\usepackage{amsbsy,amssymb}
\usepackage{wrapfig}
\usepackage{microtype}

\makeatletter
\long\def\@makecaption#1#2{\ifx\@captype\@IEEEtablestring%
	\footnotesize\begin{center}{\normalfont\footnotesize #1}\\
		{\normalfont\footnotesize\scshape #2}\end{center}%
	\@IEEEtablecaptionsepspace
	\else
	\@IEEEfigurecaptionsepspace
	\setbox\@tempboxa\hbox{\normalfont\footnotesize {#1.}~~ #2}%
	\ifdim \wd\@tempboxa >\hsize%
	\setbox\@tempboxa\hbox{\normalfont\footnotesize {#1.}~~ }%
	\parbox[t]{\hsize}{\normalfont\footnotesize \noindent\unhbox\@tempboxa#2}%
	\else
	\hbox to\hsize{\normalfont\footnotesize\hfil\box\@tempboxa\hfil}\fi\fi}
\makeatother

\usepackage{tikz,pgf}
\usetikzlibrary{patterns}
\usetikzlibrary{calc}

\usepackage{pgfplots}

\newtheorem{theorem}{Theorem}

\newtheorem{definition}{Definition}

\def\BibTeX{{\rm B\kern-.05em{\sc i\kern-.025em b}\kern-.08em
    T\kern-.1667em\lower.7ex\hbox{E}\kern-.125emX}}

\begin{document}

\title{Virtual Cell Clustering with Optimal Resource
	Allocation  to Maximize Capacity
}

\author{Michal Yemini and Andrea J. Goldsmith

\thanks{This research was supported by AFOSR Grant FA9550-12-1-0215, ONR Grant  
	N000141512527, and the Center for Science of Information under Grant CCF-0939370.}
\thanks{The authors are with the Department of Electrical Engineering, Stanford
	University, Stanford, CA, 94305 USA.}}

\maketitle

\begin{abstract}
This work proposes a new resource allocation optimization and network management framework for wireless networks using neighborhood-based optimization rather than fully centralized or fully decentralized methods. We propose hierarchical clustering with a minimax linkage criterion for the formation of the virtual cells. Once the virtual cells are formed, we consider two cooperation models: the interference coordination model and the coordinated multi-point decoding model. In the first model base stations in a virtual cell decode their signals independently, but allocate the communication resources cooperatively. In the second model base stations in the same virtual cell allocate the communication resources and decode their signals cooperatively. We address the resource allocation problem for each of these cooperation models. For the interference coordination model this problem is an NP-hard mixed-integer optimization problem whereas for the coordinated multi-point decoding model it is convex. Our numerical results indicate that proper design of the neighborhood-based optimization leads to significant gains in sum rate over fully decentralized optimization, yet may also have a significant sum rate penalty compared to fully centralized optimization. In particular, neighborhood-based optimization has a significant sum rate penalty compared to fully centralized optimization in the coordinated multi-point model, but not the interference coordination model.
\end{abstract}


\section{Introduction}

The demand for increased capacity in cellular networks continues to grow, which is  driving the deployment of spectrally-efficient
 small cells \cite{4623708,6768783,6171992,anpalagan_bennis_vannithamby_2015}. While the deployment of small cells leads to significant capacity gains over macrocell-only systems, the proximity of small cell base stations (BSs) to one another can cause severe interference between them. This interference must be managed carefully to maximize the overall network capacity. Thus, powerful interference mitigation methods as well as optimal resource allocation schemes  that involve multiple cells must be developed for 5G  networks. 

In this work we investigate a flexible network  structure for cellular systems where, instead of each BS serving all users within its own cell independently, several BSs act cooperatively to create a “virtual cell” with joint resource allocation.
 In order to design cellular networks that are composed of virtual cells, we address in this work the following two design challenges:  
  1) Creating the virtual cells, i.e., clustering the BSs and users into virtual cells. 2) Allocating the resources in each virtual cell.
In this work we address the uplink resource allocation problem for joint channel allocation and power allocation for the single user detection scenario. We also address the resource allocation problem for coordinated multi-point decoding scenarios in which BSs in a virtual cell jointly decode the signals that they receive.

BS and user clustering as part of a resource allocation strategy is discussed in the Cooperative Multi-Point (CoMP) literature, see for example  \cite{6530435,6707857,6181826,6555174,5594575,5502468,6655533,4533793,5285181,6786390,8260866}. The work \cite{7839266} presents an extensive literature survey of cell clustering for CoMP in wireless networks. The clustering of BSs and users can be divided into three groups: 1) Static clustering which considers a cellular network whose cells are clustered statically. Hence, the clustering does not adapt to network changes. Examples for static clustering algorithms are presented in \cite{6530435,6181826,6707857,6555174}. 2) Semi-dynamic clustering, in which static clusters are formed but the cluster affiliation of users is adapted according to the networks changes. Examples for such algorithms are presented in  \cite{5594575,5502468,6655533}. 3) Dynamic clustering in which the clustering of both BSs and users adapts to changes in the network. Examples for dynamic clustering algorithms are presented in \cite{4533793,5285181,6786390,8260866}. 

Resource allocation in virtual cells is  closely related to cloud radio access networks  \cite{5594708,CIT-048,6924850,7487951,6601765} in which several cells act cooperatively. The coordination between the cells can be divided into the following categories: 1) Interference coordination in which only channel states are available at the coordinated BSs. 2) Full cooperation in which BSs share not only channel states but also the data signals they receive. 3) Rate limited coordination in which the BSs exchange data via a limited-capacity backhaul. 4) Relay-assisted cooperation in which cooperation is carried out by dedicated relay nodes that connect users from different cells and BSs. 
In addition, resource allocation in virtual cells is also closely related to the interference mitigation paradigm  called Cooperative Multi-Point (CoMP) (see \cite{5706317}) that encompasses several cooperation models. Two such models are the Uplink Interference Prediction model in which cooperation is allowed in the resource allocation stage only, and the Uplink Joint Detection model that allows BS cooperation in both the resource allocation and decoding stages. 

In this work we investigate a flexible cooperative resource allocation structure for cellular systems  where, instead of each BS serving all users within its own cell independently, several BSs act cooperatively to create a “virtual cell”. We consider two BS cooperation models for the uplink communication in virtual cells. The first model allows for cooperation in the resource allocation stage only, whereas the second model allows for cooperation in both the resource allocation and the decoding stages. We  refer to the first model as the interference coordination model and to the second as the coordinated multi-point model.
Our work  \cite{YeminiGoldsmith2} considers the coordinated multi-point decoding model in which  BSs jointly decode their messages assuming infinite capacity backhaul links between BSs in the same virtual cell. Additionally, in
\cite{YeminiGoldsmith1} we propose channel and power allocation schemes for the interference coordination model. 
 This manuscript presents a unified framework that evaluates  both cooperation models analyzed in \cite{YeminiGoldsmith2} and \cite{YeminiGoldsmith1}. It  extends the analysis of the resource allocation schemes presented in \cite{YeminiGoldsmith1}, and also further evaluates and compares the network  optimization schemes presented in both \cite{YeminiGoldsmith2} and \cite{YeminiGoldsmith1}. 

Clustering as part of a resource allocation strategy in wireless networks is also investigated in the ultra-dense networks literature, see for example \cite{7008373,7579583,7794900,6786390,7248710,8110665,8496818}. 
These works can be categorized into two groups:
cell clustering (see \cite{7008373,7579583,7794900}), in which the existing cells of a cellular networks are merged, and
user-centric clustering (see \cite{6786390,7248710,8110665}), in which each user chooses a subset of BSs to communicate with. The work presented in this manuscript differs from these works in several key aspects. 
First, our channel state information model differs from that of the aforementioned works which either assume that the inter-cluster interference is perfectly known for all the channels in the network \cite{7008373,7579583,7794900,8496818}, or strictly statistical  for all the channels in the network \cite{7248710,8110665,6786390}. In our setup we assume perfect channel state information inside each virtual cell but no channel information regarding users in different virtual cells. We note that our resource allocation schemes can be  adapted to statistical knowledge regarding the inter-cluster interference.  
Second, in addition to proposing a clustering scheme to create virtual cells, we also address both the channel and power allocation problems.    
In contrast, the analysis presented in the aforementioned works are limited to the channel allocation problem and do not address the power allocation problem within the clusters. Instead it is assumed that the power allocation is fixed. A fixed power allocation can degrade significantly the performance of cooperative models, such as the coordinated multi-point decoding model, in which BSs jointly decode
the signals that they receive. 
Additionally, to the best of our knowledge, prior works optimizing performance based on cell clustering or CoMP did not consider how performance varied with the number of clusters or with the user affiliation rules.

Our work is also related to the concept of Software Defined Networks (SDN), introduced in \cite{6994333,6739370,7000974,1237143,7473831}. The underlying idea behind SDN is the separation of the data plane, which carries the data in the network, and the control plane, which determines how packets in the network are forwarded. Theoretically, the concept of SDN can be harnessed in limiting the interference in the network by allocating the resources in the network centrally \cite{6385040,6385039}. However,  the very thing that makes SDN’s centralized control plane attractive also renders its implementation complexity challenging due to the required flexibility. These complexity issues are more severe in wireless communication networks employing SDN because of their time-varying nature, which requires fast updating rules for the control plane. Creating virtual cells that are composed of several cells can assist in managing wireless network and close the gap between the promising concept of SDN and the difficulties that arise in its implementation.

\subsection{Main Contributions:}
This work extends the concept of cellular networks while preserving several of its key desirable properties, such as simple user association rules and dividing the network into independent cells that may cooperate to suppress interference. We call this network paradigm a cellular network with virtual cells. 

A cellular network design with virtual cells has the following benefits: 
\begin{enumerate}
	\item  improves network performance while balancing the computational complexity of optimal resource allocation
	\item  uses both local and global network information
	\item ensures that local changes in the network do not cause  a ``butterfly effect" in which the allocation of resources across the whole network design must be recalculated due to a local change.
\end{enumerate} 
We create the virtual cells by clustering the BSs, instead of users, in the network, and then associate users with the clustered BSs. We cluster BSs based on the hierarchical clustering method with minimax linkage criterion that creates a dendrogram. The dendrogram shows which clusters are merged when  the number of clusters is decreased and which are separated when this number is increased.  We propose using this clustering approach since it enjoys the unique property that decreasing or increasing the number of clusters  affects only the clusters that are being merged or separated, while leaving all others unchanged. By contrast, in other clustering methods, such as K-means or spectral clustering, even a small variation in the number of clusters requires the reclustering of the whole network, which may cause a global change. This is undesirable behavior for wireless communication networks since  the channel state information between all users in the new virtual cells and the new virtual BSs must be estimated.
 Thus, we propose using hierarchical clustering in which the number of clusters can adapt efficiently to the current state of the network without requiring an overall update in the network. Additionally, the method we propose requires only local channel state information that is used in the user association rule and in computing the resource allocation scheme inside the virtual cells. The BS clustering which constructs the ``backbone'' of the network does not require knowledge of the channel state between all the users and BSs in the network.

To optimize the performance of cellular networks with virtual cells we also develop resource allocation schemes for virtual cells in the single user detection scenario, and compare them to previously proposed resource allocation schemes for heterogeneous cells. Interestingly, numerical results show that the  performance of these resource allocation schemes depends on the number  of virtual cells in the network. Additionally, we address resource allocation for the coordinated multi-point decoding scenario. The resource allocation in both setups uses local channel state information, that is, we assume that the BSs in a virtual cell acquire the channel state information between them and all the users in the virtual cells.  Finally we note that, while we do not suppress interference between virtual cells in the resource allocation stage, as we decrease the number of virtual cells, interference is dominated by interference within the virtual cell so that our resource allocation scheme mitigates this dominant interference. 

\subsection{Outline and Notation}
The remainder of this paper is organized as follows. Section \ref{sec:problem_formualtion} presents the problem formulation that we analyze in this work. Section \ref{sec:virtual_cell_create} describes the method for forming the virtual cells. Sections \ref{sec:joint_power_allocation} and \ref{sec:alternating_optimization} present several algorithms for allocating resources  in the interference coordination model.
In particular, Section \ref{sec:joint_power_allocation} proposes a  joint channel  and power allocation scheme. Section  \ref{sec:alternating_optimization} proposes  channel  and power allocation algorithms based on an alternating optimization in which the resource allocation is calculated by alternating between a channel  and  power allocation problem.  Section \ref{sec:alternating_optimization}  presents three channel allocation schemes that we evaluate: a user-centric one that we propose and two existing ones,  a BS centric scheme and a sum rate maximization matching scheme. Section \ref{sec:joint_decoding} presents an optimal resource allocation scheme in virtual cells for the coordinated multi-point decoding model. Section \ref{se:simulation} presents numerical results of the average system sum rate for all of our proposed clustering and resource allocation methods. Finally, \ref{sec:conclusion} summarizes and concludes this work.

\textit{Notation:} The following notations are used throughout this paper. Vectors are denoted by  boldface lowercase letters whereas matrices are denoted by  boldface uppercase letters. We denote the transpose of a vector $\boldsymbol a$ by $\boldsymbol a'$, and the conjugate transpose of a matrix $\boldsymbol A$ by $\boldsymbol A^{\dagger}$. The expected value of a random variable $x$ is denoted by $E(x)$. Additionally, we denote the covariance matrix of a random vector $\boldsymbol x$ by 
$\text{cov}(\boldsymbol x)$.  $\det(\boldsymbol A)$ denotes the determinant of a square matrix $\boldsymbol A$. Finally, $\mathbbm{1}_{\mathcal{E}}$ denotes the indicator function; it is equal to one if the event $\mathcal{E}$ is true and zero otherwise. Finally the cardinality of a set $\mathcal{S}$ is denoted by $|\mathcal{S}|$.
 
\section{Problem Formulation}\label{sec:problem_formualtion}
We consider  a communication network that comprises a set of BSs (BSs) $\mathcal{B}$, a set of users $\mathcal{U}$ and a set of frequency bands $\mathcal{K}$. The users communicate with their BSs and these transmissions interfere with one another. Each user $u\in\mathcal{U}$ has a maximal transmission power of $\overline{P}_u$ dBm.
The BSs and users are clustered into virtual cells that must fulfill the following characteristics.

\subsection{Virtual Cells}\label{sec:virtual_cell_requirements}

\begin{definition}[Virtual BS]
Let $b_1,..,b_n$ be $n$ BSs in the set of BSs $\mathcal{B}$, we call the set $\{b_1,..,b_n\}$ a virtual BS.
\end{definition}
\begin{definition}[Proper clustering]	
Let $\mathcal{B}$ be a set of BSs,  $\mathcal{U}$ be a set of users. Denote   $\mathcal{V}=\{1,\ldots,V\}$.
For every $v$, define the sets $\mathcal{B}_v\subset \mathcal{B}$ and $\mathcal{U}_v\subset \mathcal{U}$ .
We say that the set $\mathcal{V}$ is a proper clustering of the sets  $\mathcal{B}$ and  $\mathcal{U}$  if $\mathcal{B}_v$ is a partition of the sets $\mathcal{B}$ and $\mathcal{U}$. That is,
$\bigcup_{v\in\mathcal{V}}\mathcal{B}_v = \mathcal{B}$, $\bigcup_{v\in\mathcal{U}}\mathcal{U}_v = \mathcal{U}$. Additionally,
  $\mathcal{B}_{v_1}\cap\mathcal{B}_{v_2}=\emptyset$ and $\mathcal{U}_{v_1}\cap\mathcal{U}_{v_2}=\emptyset$ for all $v_1,v_2\in\mathcal{V}$ such that $v_1\neq v_2$.
\end{definition}

\begin{definition}[Virtual cell]
Let $\mathcal{B}$ be a set of BSs, $\mathcal{U}$ be a set of users, and  $\mathcal{V}$ be a proper clustering of $\mathcal{B}$ and $\mathcal{U}$. For every $v\in\mathcal{V}$  the virtual cell  $\mathcal{C}_v$ is composed of the virtual BS $\mathcal{B}_v$ and the set of users $\mathcal{U}_v$.	
\end{definition}

This condition ensures that every BS and every user belongs to exactly one virtual cell. This implies that all the transmission power of a user is dedicated to communicating with BSs in the same virtual cell, thus power allocation can be optimized in a virtual cell.    

Let $\mathcal{V}$ be a proper clustering of the set of BSs $\mathcal{B}$ and the set of users $\mathcal{U}$, and let   $\{\mathcal{C}_v\}_{v\in\mathcal{V}}$ be the set of virtual cells that $\mathcal{V}$ creates.
In each virtual $\mathcal{C}_v$ we assume that the BSs that compose the virtual BS $\mathcal{B}_v$ jointly allocate their resources.

\subsection{The Uplink Resource Allocation Problem  for the Interference Coordination Model}\label{subsection:uplink_interference_coordination_problem}

In each virtual cell we consider the uplink resource allocation problem in which  all the BSs in the virtual cell jointly optimize the channel allocation and the transmission power  of the users within the virtual cell. Further, we  consider  single user detection in which every BS $b$ decodes each of its codewords separately.
That is, suppose that users $u_1$ and $u_2$ are both served by BS $b$, then $b$ decodes the codeword of $u_1$ treating the codeword of $u_2$ as noise, and decodes the codeword of $u_2$ treating the codeword of $u_1$ as noise. We refer to this model as the  interference coordination model.

While each user can communicate with all the BSs in its virtual cell, it follows by \cite{1237143} that, given a power allocation scheme, the maximal communication rate for each user is achieved when the message is decoded by the BS with the highest SINR for this user.
Recall that $\mathcal{K}$ is the set of frequency bands.
Denote by $h_{u,b,k}$  the channel coefficient of the channel from user $u\in\mathcal{U}$ to BS $b$ over frequency band $k$, and let $P_{u,k}$ be the transmit power of user $u$ over frequency band $k$. Further, let $\sigma^2_{b,k}$ denote the noise power at BS $b$ over frequency band $k$, and let $W_k$ denote the bandwidth of band $k$.
The uplink resource allocation problem in each virtual cell $\mathcal{C}_v$, ignoring  interference from other virtual cells, is given by:
\begin{flalign}\label{eq:no_decoding_cooperation_single_discrete}
\max & \sum_{b\in\mathcal{B}_v}\sum_{u\in\mathcal{U}_v}\sum_{k\in\mathcal{K}}
\gamma_{u,b,k}W_k\log_2\left(1+\frac{|h_{u,b,k}|^2P_{u,k}}{\sigma^2_{b,k}+J_{u,b,k}}\right)\nonumber\\
\text{s.t.: } &  0\leq P_{u,k},\quad \sum_{k\in\mathcal{K}}P_{u,k} \leq \overline{P}_u,\quad \forall\: u\in \mathcal{U}_v,k\in\mathcal{K},\nonumber\\
&\hspace{-0.15cm} \sum_{\substack{\tilde{u}\in\mathcal{U}_v,\\ \tilde{u}\neq u}} |h_{\tilde{u},b,k}|^2P_{\tilde{u},k}=  J_{u,b,k},\: \forall  u\in\mathcal{U}_v,b\in \mathcal{B}_v,k\in\mathcal{K} \nonumber\\
&\gamma_{u,b,k}\in\{0,1\},\quad \sum_{b\in\mathcal{B}_v}\gamma_{u,b,k}\leq 1,\quad \forall\:u\in \mathcal{U}_v,b\in \mathcal{B}_v,k\in\mathcal{K}.
\end{flalign}
This is a mixed-integer programming problem that is NP-hard. Sections \ref{sec:joint_power_allocation} and \ref{sec:alternating_optimization} present two different approaches to approximate this problem for a given virtual cell. The first approach, presented in Section \ref{sec:joint_power_allocation},  translates this problem from a mixed-integer programming problem to an equivalent problem with continuous variables. The second approach, presented in Section \ref{sec:alternating_optimization},  
approximates the optimal solution by  solving a user-centric channel allocation problem, and a power allocation problem, alternately.

\subsection{The Uplink Resource Allocation Problem for Coordinated Multi-Point Decoding}\label{subsection:uplink_joint_decoding_problem}
In the coordinated multi-point decoding model BSs jointly decode the signals that they receive. This model can be realized, for example, based on cloud decoding of the signals received by all BSs under the assumption that the BS communication to the cloud has unconstrained capacity. This model is equivalent to a multiple access channel (MAC) with a single transmitting antenna at each user and  multiple  antennas corresponding to all BS antennas at the receiver. Recalling that $\mathcal{K}$ is the set of frequency bands,  denote by $x_{u},k$ the signal of user $u$ on frequency band $k$, and by $y_{b,k}$  the received signal at BS $b$ for band $k\in\mathcal{K}$. For the sake of clarity, we label the BSs in the cluster $v$ by $b_1,\ldots, b_{|\mathcal{B}_v|}$, and label the users in cluster $v$ by  $u_1,\ldots,u_{|\mathcal{U}_v|}$.  
Denote  $\boldsymbol y_{v,k}\triangleq(y_{b_1,k},\ldots,y_{b_{|\mathcal{B}_v|},k})'$ and let $\boldsymbol x_{v,k}\triangleq(x_{u_1,k},\ldots,x_{u_{|\mathcal{U}_v|},k})'$.  The receiving signal at BS $b\in \mathcal{B}_v$, ignoring the interference from other clusters, in frequency band $k$ is
\begin{flalign}
y_{b,k} = \sum_{i=1}^{|\mathcal{U}_v|}h_{u_i,b,k} x_{u_i,k}+n_{b,k},
\end{flalign}
where $h_{u_i,b,k}$ is the channel coefficient from user $u_i$ in $v$ to the BS $b$ in $v$ over frequency band $k$, and $n_{b,k}$ is a white Gaussian noise at BS $b$ over frequency band $k$. 

Let $\boldsymbol h_{u_i,k} = (h_{u_1,b_1,k},\ldots,h_{u_i,b_{|\mathcal{B}_v|},k})'$ be the channel coefficient vector between user $u_i$ in $v$ to all the BSs in cluster $v$. Then the receiving signal vectors at the BSs in $v$ are
\begin{flalign}
\boldsymbol y_{v,k} &= \sum_{i=1}^{|\mathcal{U}_v|}\boldsymbol h_{u_i,k} x_{u_i,k}+\boldsymbol n_{v,k},
\end{flalign}
where 
$\boldsymbol n_{v,k}=(n_{b_1,k},\ldots,n_{b_{|\mathcal{B}_v|,k}})$ is a white noise vector at the BSs.

 Let $\boldsymbol C_{v,k}=\text{cov}\left(\boldsymbol{x}_{v,k}\right)$ and $\boldsymbol N_{v,k} = \text{cov}(\boldsymbol n_{v,k})$; the sum capacity of the uplink in the virtual cell is then:
\begin{flalign}\label{eq:uplink_problem_clean}
\max &\sum_{k\in\mathcal{K}}W_k\log_2\det\left(\boldsymbol I+\sum_{u\in\mathcal{U}_v}p_{u,k}\boldsymbol h_{u,k} \boldsymbol h_{u,k}^{\dagger}\boldsymbol{N}_{v,k}^{-1}\right)\nonumber\\
\text{s.t.: } & \sum_{k\in\mathcal{K}} p_{u,k}\leq \overline{P}_u,\quad p_{u,k}\geq 0.
\end{flalign}

We note that while interference between virtual cells is not addressed in this work, as the number of virtual cells is decreased, each virtual cell becomes
 larger, and the interference inside the virtual cells becomes the dominant interference.  This interference is mitigated in  (\ref{eq:no_decoding_cooperation_single_discrete}) and (\ref{eq:uplink_problem_clean}) to improve network performance. Additionally, we note that if an approximated inter-cluster interference is known to be $i_{b,k}$ at BS $b$ at frequency band $k$, then  term  $\sigma_{b,k}^2$ can be replaced with $\sigma_{b,k}^2+i_{b,k}$ in the interference coordination model. Similarly, in coordinated multi-point decoding, the noise covariance matrix $\boldsymbol N_{v,k}$ can be replaced with  the term  $\boldsymbol N_{v,k}+\boldsymbol I_{v,k}$ where $\boldsymbol I_{v,k}$ is some approximation for the covariance matrix of inter-cluster interference in the virtual cell $v$.  

\section{Forming the Virtual Cells}\label{sec:virtual_cell_create}
This section presents the clustering approach that creates the virtual cells within which the resource allocation scheme we present in Sections \ref{sec:joint_power_allocation}-\ref{sec:joint_decoding} operate.

\subsection{Base Station Clustering via Hierarchical Clustering with Minimax Linkage Criterion}
A hierarchical clustering algorithm creates a linkage tree, using a linkage criterion, that shows which clusters are merged when the number of clusters is decreased, and which are separated when this number is increased. This linkage tree is called a dendrogram.  
We propose using  the hierarchical clustering algorithm to cluster BSs, since it enjoys the unique property that decreasing or increasing the number of clusters  only affects the clusters that are being merged or separated. Thus, the number of clusters can adapt efficiently to the current state of the network without requiring a full clustering update. By contrast, in other clustering methods, such as K-means or spectral clustering, even a small variation in the number of clusters requires a full clustering update. This is undesirable in wireless networks since a large setup time and overhead for each reclustering is needed for information acquisition and other message passing.

Furthermore, we propose using the  hierarchical clustering algorithm with the minimax linkage criterion proposed in \cite{BienTibshirani2011} and that we depict in Algorithm \ref{algo:hierarchical_clustering}. This algorithm gets  a set of points $S$ and produces the clusterings $B_1,\ldots,B_n$, where $B_m$ is the clustering of size $m$. The algorithm defines the center of a cluster to be the member of the cluster with the minimal maximal distance to all other members in the cluster. This minimal maximal distance is the cluster radius. Then, in every step, the minimax linkage criterion merges the two clusters that will jointly have the smallest radius out of all merging possibilities.
Since interference tends to increase on average as the distance between interferers is decreased, at each stage the minimax linkage criterion  merges the two clusters of BSs that maximize the smallest anticipated interference at the center of the new cluster caused by the cluster BSs.  In addition, the minimax linkage criterion benefits from fulfilling several desirable properties in cluster analysis, as discussed in  \cite{BienTibshirani2011}, that other linkage criteria such as the centroid linkage criteria do not fulfill. Next, we formally depict the hierarchical clustering algorithm with minimax linkage criterion.

Let $d:\mathbb{R}^2\times\mathbb{R}^2\rightarrow\mathbb{R}$ be the Euclidean distance function, and let $S$ be a set of points in $\mathbb{R}^2$. We then define the following:
\begin{definition}[Radius of a set around point]
	The radius of $S$ around $s_i \in S$  is defined as $r(s_i,S)=\max_{s_j\in S}\:d(s_i,s_j)$.
\end{definition}
\begin{definition}[Minimax radius]
	The minimax radius of $S$ is defined as $r(S) = \min_{s_i\in S}\: r(s_i,S)$.
\end{definition}
\begin{definition}[Minimax linkage]
	The minimax linkage between two sets of points $S_1$ and $S_2$ in $\mathbb{R}^2$ is defined as $d(S_1,S_2) = r(S_1\cup S_2)$.
\end{definition}

Let $S=\{s_1,\ldots,s_n\}$ be the set of locations of the BSs in $\mathcal{B}$. We use Algorithm \ref{algo:hierarchical_clustering} below with the input $S$ to create the  virtual BSs for each number of clusters $m$. This produces the dendrogram which shows what clusters are merged as the number of clusters is decreased.

\setlength{\textfloatsep}{.7cm}
\begin{algorithm}
	\caption{}\label{algo:hierarchical_clustering}
	\begin{algorithmic}[1]		
		\State Input: A set of point $S=\{s_1,\ldots,s_n\}$;
		\State Set $B_n = \left\{\{s_1\},\dots,\{s_n\}\right\}$;
		\State Set $d(\{s_i\},\{s_j\})=d(s_i,s_j),\:\forall s_i,s_j\in S$;	
		\For {$m = n-1,\ldots,1$}
		\State Find $(S_1,S_2) = \arg\min_{\stackrel{G,H\in B_{m+1}:}{G\neq H}} d(G,H)$;
		\State  Update $B_{m} = B_{m+1} \bigcup \{S_1\cup S_2\} \setminus \{S_1,S_2\}$;
		\State Calculate $d(S_1\cup S_2,G)$ for all $G\in B_m$;
		\EndFor
	\end{algorithmic}
	
\end{algorithm}

\subsection{Users' Affiliation with Clusters}\label{sec_user_affil}
To create the virtual cells, we consider two affiliation rules: 
\begin{enumerate}
	\item Closest BS rule in which each user is affiliated with its closest BS.  
	\item Best channel rule in which  each user is affiliated with the BS to which it has the best channel  (absolute value of the channel coefficient). 
\end{enumerate}
	Then each user is associated with the virtual BS that its affiliated BS is part of.
This way every virtual BS and it associated users compose a virtual cell.
It is easy to verify that the formation of the virtual cells we propose fulfills the requirement presented in Section \ref{sec:virtual_cell_requirements}.

 The combination of creating virtual cells by using global network information for BS clustering and local network information to associate users with virtual cells creates an easy-to-manage network architecture that does not require a global update when local changes in the network occur. 

\section{Channel  and Power  Allocation for the Interference Coordination Model}\label{sec:joint_power_allocation}
This section introduces the first  resource allocation scheme we propose for the interference coordination model. This scheme is found by converting the problem (\ref{eq:no_decoding_cooperation_single_discrete}) to an equivalent continuous variable problem and then solving the new problem via a convex approximation.
\subsection{An Equivalent Continuous Variable Resource Allocation Problem}
We can represent the problem (\ref{eq:no_decoding_cooperation_single_discrete}) by an equivalent problem with continuous variables. Suppose that, instead of sending a message to at most one single BS at each frequency band, a user sends messages to all BSs. The signal of user $u\in\mathcal{U}_v$ over frequency band $k$ is then given by $x_{u,k}=\sum_{b\in\mathcal{B}_v}x_{u,b,k}$ where $x_{u,b,k}$ is the part of the signal of user $u$ that is transmitted over frequency band $k$ and is  intended to be decoded by BS $b$. Let $P_{u,b,k}$ be the power allocation of the part of the signal of user $u$ that is transmitted over frequency band $k$ and is  intended to be decoded by BS $b$.; i.e. $P_{u,b,k}=E\left( x_{u,b,k}^2\right)$, where $E\left( x_{u,b,k}^2\right)$ denotes the expected value of $x_{u,b,k}^2$.
We next prove that (\ref{eq:no_decoding_cooperation_single_discrete}) can in fact be written in the following equivalent form:
\begin{flalign}\label{eq:no_decoding_cooperation_single_continuous}
\max & \sum_{b\in\mathcal{B}_v}\sum_{u\in\mathcal{U}_v}\sum_{k\in\mathcal{K}}
W_k\log_2\left(1+\frac{|h_{u,b,k}|^2P_{u,b,k}}{\sigma^2_{b,k}+J_{u,b,k}}\right)\nonumber\\
\text{s.t.: } & 0\leq P_{u,b,k},\quad  \sum_{b\in\mathcal{B}_v}\sum_{k\in\mathcal{K}} P_{u,b,k}\leq \overline{P}_u,\quad \forall \: u\in\mathcal{U}_v,b\in\mathcal{B}_v,k\in\mathcal{K},\nonumber\\
& \hspace{-0.55cm}\sum_{\substack{(\tilde{u},\tilde{b})\in\mathcal{U}_v\times \mathcal{B}_v,\\(\tilde{u},\tilde{b})\neq (u,b)}}\hspace{-0.5cm} |h_{\tilde{u},b}|^2P_{\tilde{u},\tilde{b},k}=  J_{u,b,k},\: \forall\:  u\in\mathcal{U}_v,b\in \mathcal{B}_v,k\in\mathcal{K}.
\end{flalign}

\begin{theorem}\label{theorem:equivalence:discrete_continuous}
The mixed-integer programming problem (\ref{eq:no_decoding_cooperation_single_discrete}) and the continuous variables problem (\ref{eq:no_decoding_cooperation_single_continuous})  are equivalent. 	
\end{theorem}

\begin{IEEEproof}
	The equivalence of (\ref{eq:no_decoding_cooperation_single_discrete}) and (\ref{eq:no_decoding_cooperation_single_continuous}) is argued as follows.
	
	First, the solution of (\ref{eq:no_decoding_cooperation_single_discrete}) can be achieved by the solution of (\ref{eq:no_decoding_cooperation_single_continuous}) by setting $x_{u,b,k}=0$ whenever $\gamma_{u,b,k}=0$, and $E \left(x_{u,b,k}^2\right) = P_{u,k}$ whenever $\gamma_{u,b,k}=1$. Thus the maximal sum rate that is found by solving (\ref{eq:no_decoding_cooperation_single_continuous}) upper bounds the maximal sum rate  that is found by solving (\ref{eq:no_decoding_cooperation_single_discrete}).
	On the other hand, suppose that the optimal transmission power of user $u$ using frequency band $k$, given the transmission power of all other users, is $P_{u,k}$, that is $P_{u,k} = \sum_{b\in\mathcal{B}_v}P_{u,b,k}$.
	It follows by the duality between the multiple-access channel and the broadcast channel that is proved in  \cite{1237143} that the optimal power allocation $(P_{u,b,k})_{b\in\mathcal{B}_v}$ for user $u$ in frequency band $k$, given the power allocation of all other users, is to allocate all its transmission power $P_{u,k}$ over frequency band $k$ to the transmission to the BS with the highest SINR.
	It follows that the maximal sum rate of (\ref{eq:no_decoding_cooperation_single_continuous}) cannot be larger than that of (\ref{eq:no_decoding_cooperation_single_discrete}). Thus, the two problems (\ref{eq:no_decoding_cooperation_single_discrete}) and (\ref{eq:no_decoding_cooperation_single_continuous}) are equivalent.
\end{IEEEproof}

\subsection{Solving an Approximation of the Continuous Variable Resource Allocation Problem Optimally}\label{sec:continuous_HSINR_gradient}
In the following, we solve  problem (\ref{eq:no_decoding_cooperation_single_continuous}).
Denote:
\begin{flalign}\label{eq:SINR_def}
\text{SINR}_{u,b,k}(\boldsymbol P) =\frac{|h_{u,b,k}|^2 P_{u,b,k}}{\sigma^2_b+\sum_{\substack{(\tilde{u},\tilde{b})\in\mathcal{U}_v\times \mathcal{B}_v,\\(\tilde{u},\tilde{b})\neq (u,b)}} |h_{\tilde{u},b,k}|^2P_{\tilde{u},\tilde{b},k}},
\end{flalign}
where $\boldsymbol P = (P_{u,b,k})_{(u,b,k)\in\mathcal{U}_{v}\times\mathcal{B}_{v}\times\mathcal{K}}$ is the matrix of the transmission power.

Using the high SINR approximation \cite{5165179}
\begin{flalign}\label{eq:high_SINR_approx_improved}
\log(1+z)\geq \alpha(z_0)\log z+\beta(z_0),
\end{flalign}
where
\begin{flalign}\label{eq:alpha_beta_def}
\alpha(z_0) = \frac{z_0}{1+z_0},\qquad\beta(z_0) =\log(1+z_0)-\frac{z_0}{1+z_0}\log{z_0},
\end{flalign}
we obtain the approximated iterative problem (\ref{eq:iterative_alpha_approx})
where $\alpha_{u,b,k}^{(m)}=\alpha(\text{SINR}_{u,b,k}(\boldsymbol P^{(m-1)}))$, $\beta_{u,b,k}^{(m)}=\beta(\text{SINR}_{u,b,k}(\boldsymbol P^{(m-1)}))$ and $\alpha_{u,b,k}^{(0)}=1$, $\beta_{u,b,k}^{(0)}=0$ for all $u\in\mathcal{U}_v$, $b\in\mathcal{B}_v$ and $k\in\mathcal{K}$.

\begin{flalign}\label{eq:iterative_alpha_approx}
\boldsymbol P^{(m)} =& \arg\max_{\boldsymbol P} \sum_{b\in\mathcal{B}_v}\sum_{u\in\mathcal{U}_v}\sum_{k\in\mathcal{K}}
W_k\left[\alpha_{u,b,k}^{(m)}\log_2\left(\frac{|h_{u,b,k}|^2P_{u,b,k}}{\sigma^2_{b,k}+
	\sum_{\substack{(\tilde{u},\tilde{b})\in\mathcal{U}_v\times \mathcal{B}_v,\\(\tilde{u},\tilde{b})\neq (u,b)}} |h_{\tilde{u},b,k}|^2P_{\tilde{u},\tilde{b},k}}\right)+\beta_{u,b,k}^{(m)}\right]\nonumber\\
&\text{s.t.: } \: 0\leq P_{u,b,k},\quad   \sum_{b\in\mathcal{B}_v}\sum_{k\in\mathcal{K}} P_{u,b,k}\leq \overline{P}_u,\quad \forall \: u\in\mathcal{U}_v,b\in\mathcal{B}_v,k\in\mathcal{K}\nonumber\\
& \sum_{\substack{(\tilde{u},\tilde{b})\in\mathcal{U}_v\times \mathcal{B}_v,\\(\tilde{u},\tilde{b})\neq (u,b)}} |h_{\tilde{u},b,k}|^2P_{\tilde{u},\tilde{b},k}=  J_{u,b,k},\quad \forall \: u\in\mathcal{U}_v,b\in \mathcal{B}_v,k\in\mathcal{K}.
\end{flalign}

It is left to solve the problem (\ref{eq:iterative_alpha_approx}). By transforming the variables of the problem using $P_{u,b,k}=\exp(g_{u,b,k})$ and noticing that the terms $\beta_{u,b,k}^{(m)}$ do not affect the optimal power allocation, we get the equivalent convex problem:
\begin{flalign}\label{sol_continuous_power_approx}
&\ln(\boldsymbol P^{(m)}) = \arg\max \sum_{b\in\mathcal{B}_v}\sum_{u\in\mathcal{U}_v}\sum_{k\in\mathcal{K}}
W_k\alpha_{u,b,k}^{(m)}\cdot\log_2\left(\frac{|h_{u,b,k}|^2\exp(g_{u,b,k})}{\sigma^2_{b,k}+
	\sum_{\substack{(\tilde{u},\tilde{b})\in\mathcal{U}_v\times \mathcal{B}_v,\\(\tilde{u},\tilde{b})\neq (u,b)}} |h_{\tilde{u},b,k}|^2\exp(g_{\tilde{u},\tilde{b},k})}\right)\nonumber\\
&\text{s.t.: }    \sum_{b\in\mathcal{B}_v}\sum_{k\in\mathcal{K}} \exp(g_{u,b,k})\leq \overline{P}_u,\quad \forall\: u\in \mathcal{U}_v.
\end{flalign}

The Lagrangian of (\ref{sol_continuous_power_approx}) is given by
\begin{flalign}\label{eq:Lagrangian_dual_prob_continuous}
&L(\boldsymbol g,\boldsymbol\lambda;m) =  \sum_{b\in\mathcal{B}_v}\sum_{u\in\mathcal{U}_v}\sum_{k\in\mathcal{K}}
W_k\alpha_{u,b,k}^{(m)}\cdot\log_2\left(\frac{|h_{u,b,k}|^2\exp(g_{u,b,k})}{\sigma^2_{b,k}+
	\sum_{\substack{(\tilde{u},\tilde{b})\in\mathcal{U}_v\times \mathcal{B}_v,\\(\tilde{u},\tilde{b})\neq (u,b)}} |h_{\tilde{u},b,k}|^2\exp(g_{\tilde{u},\tilde{b},k})}\right)\nonumber\\
&\hspace{5cm}-  \sum_{u\in\mathcal{U}_v} \lambda_u\left(\sum_{b\in\mathcal{B}_v}\sum_{k\in\mathcal{K}} \exp(g_{u,b,k})-\overline{P}_u\right),
\end{flalign}
where $m$ denotes the $m$th time (\ref{sol_continuous_power_approx}) is solved.

Furthermore, the dual function of the Lagrangian is given by 
\begin{flalign}\label{eq:maximizer_Lagrangian_continuous}
q(\boldsymbol \lambda;m) = \sup_{\boldsymbol g} L(\boldsymbol g,\boldsymbol\lambda;m).
\end{flalign}
Thus the dual problem of (\ref{sol_continuous_power_approx}) is
\begin{flalign}\label{eq:prob_dual_ptob_continuous}
&\max q(\boldsymbol\lambda;m),\nonumber\\
& \text{s.t.: } \lambda_u\geq 0,\:\forall u\in\mathcal{U}_v.
\end{flalign}
Since the problem (\ref{sol_continuous_power_approx}) is convex with a non-empty interior, its duality gap is zero. Additionally, since (\ref{sol_continuous_power_approx}) has a compact domain in terms of
$P_{u,b,k}$,  it follows from \cite[Proposition 6.1.1]{Bertsekas/99} that we can solve the dual problem (\ref{eq:prob_dual_ptob_continuous}) using the gradient ascend method, that is:
\begin{flalign}\label{eq:grad_ascend}
\lambda_u^{(m,n+1)} = \left[\lambda_u^{(m,n)}+\epsilon_{\lambda}\left(\sum_{b\in\mathcal{B}_v}\sum_{k\in\mathcal{K}} \exp(g_{u,b,k}^{(m,n)})-\overline{P}_u\right)\right]^+,
\end{flalign} 
where $\boldsymbol g^{(m,n)} = (g_{u,b,k}^{(m,n)})_{u\in\mathcal{U}_v,b\in\mathcal{B}_v,k\in\mathcal{K}}$ is the maximizer of $L(\boldsymbol g,\boldsymbol\lambda^{(m,n)};m)$.

Recall that  $P_{u,b,k}=\exp(g_{u,b,k})$. It is left to solve the subproblem (\ref{eq:maximizer_Lagrangian_continuous}). Since its objective function is a strictly concave and differentiable function of $\boldsymbol g$,  
a solution is attained at the point:
\begin{flalign}\label{eq:fixed_point_prob_continuous}
&P_{u,b,k}=\frac{W_k\alpha_{u,b,k}^{(m)}}{\lambda_u\ln 2+W_k\sum_{\substack{(\tilde{u},\tilde{b})\in\mathcal{U}_v\times \mathcal{B}_v,\\(\tilde{u},\tilde{b})\neq (u,b)}}\alpha_{\tilde{u},\tilde{b},k}^{(m)}\frac{\text{SINR}_{\tilde{u},\tilde{b},k}(\boldsymbol P^{(m)})}{P_{\tilde{u},\tilde{b},k}^{(m)}|h_{\tilde{u},\tilde{b},k}|^2 }|h_{u,\tilde{b},k}|^2}.
\end{flalign}

By \cite{5165179} and \cite{414651} we can solve the fixed point (\ref{eq:fixed_point_prob_continuous})  problem iteratively:
\begin{flalign}\label{update_rule_continuous_orig}
&P_{u,b,k}^{(m,n,s+1)}=\frac{W_k\alpha_{u,b,k}^{(m)}}{\lambda_u^{(n)}\ln 2+W_k\sum_{\substack{(\tilde{u},\tilde{b})\in\mathcal{U}_v\times \mathcal{B}_v,\\(\tilde{u},\tilde{b})\neq (u,b)}}\alpha_{\tilde{u},\tilde{b},k}^{(m)}\frac{\text{SINR}_{\tilde{u},\tilde{b},k}(\boldsymbol P^{(m,n,s)})}{P_{\tilde{u},\tilde{b},k}^{(m,n,s)}|h_{\tilde{u},\tilde{b},k}|^2 }|h_{u,\tilde{b},k}|^2}
\end{flalign}
to achieve the optimal power allocation of the subproblem (\ref{eq:maximizer_Lagrangian_continuous}) where $m$ denotes the iteration number of the high SINR approximation, $n$ denotes the iteration number of the gradient ascent algorithm used to solve the dual problem, and $s$ denotes the iteration of the iterative fixed point solution. The existence of the solution is guaranteed because of the strong concavity of (\ref{eq:maximizer_Lagrangian_continuous}).

\subsection{Solving an Approximation of the Continuous Variable Resource Allocation Problem Efficiently}\label{sec:continuous_HSINR_fixed_point}

Since  the problem (\ref{sol_continuous_power_approx}) is convex with a non empty interior, its duality gap is zero, and the Karush–Kuhn–Tucker (KKT) conditions are sufficient for the points to be primal and dual optimal. The KKT conditions for (\ref{sol_continuous_power_approx}), after substituting $P_{u,b,k}=\exp(g_{u,b,k})$, are
\begin{flalign}
&P_{u,b,k}=\frac{W_k\alpha_{u,b,k}^{(m)}}{\lambda_u\ln 2+W_k\sum_{\substack{(\tilde{u},\tilde{b})\in\mathcal{U}_v\times \mathcal{B}_v,\\(\tilde{u},\tilde{b})\neq (u,b)}}\alpha_{\tilde{u},\tilde{b},k}^{(m)}\frac{\text{SINR}_{\tilde{u},\tilde{b},k}(\boldsymbol P^{(m)})}{P_{\tilde{u},\tilde{b},k}^{(m)}|h_{\tilde{u},\tilde{b},k}|^2 }|h_{u,\tilde{b},k}|^2},\quad \forall u\in\mathcal{U}_v,\\
&0=\lambda_u\left(\sum_{b\in\mathcal{B}_v}\sum_{k\in\mathcal{K}} P_{u,b,k}- \overline{P}_u\right),\quad \forall u\in\mathcal{U}_v,\\ 
& \sum_{b\in\mathcal{B}_v}\sum_{k\in\mathcal{K}} P_{u,b,k}\leq \overline{P}_u,\qquad\lambda_u\geq 0, \quad \forall u\in\mathcal{U}_v.
\end{flalign} 

Define the following iterative update rule
\begin{flalign}\label{update_rule_continuous}
&P_{u,b,k}^{(m,s+1)}=\frac{W_k\alpha_{u,b,k}^{(m)}}{\lambda_u^{(s+1)}\ln 2+W_k\sum_{\substack{(\tilde{u},\tilde{b})\in\mathcal{U}_v\times \mathcal{B}_v,\\(\tilde{u},\tilde{b})\neq (u,b)}}\alpha_{\tilde{u},\tilde{b},k}^{(m)}\frac{\text{SINR}_{\tilde{u},\tilde{b},k}(\boldsymbol P^{(m,s)})}{P_{\tilde{u},\tilde{b},k}^{(m,s)}|h_{\tilde{u},\tilde{b},k}|^2 }|h_{u,\tilde{b},k}|^2},
\end{flalign}
where $\lambda_u^{(s+1)}=0$ if
\begin{flalign}
\sum_{b\in\mathcal{B}_v}\sum_{k\in\mathcal{K}}
\frac{\alpha_{u,b,k}^{(m)}}{\sum_{\substack{(\tilde{u},\tilde{b})\in\mathcal{U}_v\times \mathcal{B}_v,\\(\tilde{u},\tilde{b})\neq (u,b)}}\alpha_{\tilde{u},\tilde{b},k}^{(m)}\frac{\text{SINR}_{\tilde{u},\tilde{b},k}(\boldsymbol P^{(m,s)})}{P_{\tilde{u},\tilde{b},k}^{(m,s)}|h_{\tilde{u},\tilde{b},k}|^2 }|h_{u,\tilde{b},k}|^2}\leq \overline{P}_u.
\end{flalign}
Otherwise $\lambda_u^{(s+1)}$ is chosen such that $\sum_{b\in\mathcal{B}_v}\sum_{k\in\mathcal{K}}P_{u,b,k}^{(m,s+1)}=\overline{P}_u$.

We have that if this update rule  converges, it must converge to a KKT point, which in turn is globally optimal. While there is no known proof that guarantees convergence, in practice convergence is  observed in simulations.

\section{Solving the Resource Allocation Problem via Alternating Optimization}\label{sec:alternating_optimization}

A more traditional approach to solving the resource allocation problem  (\ref{eq:no_decoding_cooperation_single_discrete}) separates it into two subproblems:  a channel allocation problem that sets the value of $\gamma_{u,b,k}$ to be either zero or one, and a power allocation problem that optimizes the transmission power.  Then we iteratively solve these two problems until a stopping criterion is fulfilled.
A resource allocation scheme of this type is depicted by Algorithm \ref{algo:Altenating_general}.

\setlength{\textfloatsep}{.7cm}
\begin{algorithm}
	\caption{}\label{algo:Altenating_general}
	\begin{algorithmic}[1]
		\State Notations: $\boldsymbol P^{(n)} = (P^{(n)}_{u,b,k})_{(u,b,k)\in\mathcal{U}_v\times\mathcal{B}_v\times\mathcal{K}}$, $\boldsymbol \gamma^{(n)} = (\gamma^{(n)}_{u,b,k})_{(u,b,k)\in\mathcal{U}_v\times\mathcal{B}_v\times\mathcal{K}}$;
		\State Input: $\delta>0, N_{\max}\in\mathbb{N}$;
		\State Set $n=0$,  $\delta_0 = 2\delta$;
		\State Set $P^{(0)}_{u,b,k}=\overline{P}_u/(|\mathcal{B}_v||\mathcal{K}|)$ and $\gamma^{(0)}_{u,b,k}=0$  for all $u\in\mathcal{U}_v$, $b\in\mathcal{B}_v$
		and $k\in\mathcal{K}$;
		\While{  $\delta_n>\delta$ and $n<N_{\max}$}
		\State $n=n+1$;
		\State \textbf{Channel allocation:} Given the power allocation $\boldsymbol P^{(n-1)}$,  set $\gamma^{(n)}_{u,b,k}$ to be either zero or one for every $u\in\mathcal{U}_v$, $b\in\mathcal{B}_v$ and $k\in\mathcal{K}$.		
		\State \textbf{Power allocation:}  Given  $\boldsymbol\gamma^{(n)}$, calculate $\boldsymbol P^{(n)}$
		by solving the iterative problem (\ref{sol_continuous_power_approx}) starting with some 
		initial values  $\alpha_{u,b,k}^{(0)}$, $(u,b,k)\in\mathcal{U}_v\times\mathcal{B}_v\times\mathcal{K}$.
		\State Calculate the sum rate \[R(\boldsymbol P^{(n)},\boldsymbol \gamma^{(n)})\hspace{-0.1cm} =\hspace{-0.1cm}\sum_{b\in\mathcal{B}_v}\hspace{-0.05cm}\sum_{u\in\mathcal{U}_v}\hspace{-0.05cm}\sum_{k\in\mathcal{K}}
		\gamma^{(n)}_{u,b,k}W_k\log_2\left(1+\frac{|h_{u,b,k}|^2P^{(n)}_{u,b,k}}{\sigma^2_{b,k}+J^{(n)}_{u,b,k}}\right); \]
		\State Calculate $\delta_n = R(\boldsymbol P^{(n)},\boldsymbol \gamma^{(n)})-R(\boldsymbol P^{(n-1)},\boldsymbol \gamma^{(n-1)})$;
		\EndWhile
	\end{algorithmic}
\end{algorithm}

For the sake of depicting the channel allocation schemes and the initial values of $\alpha^{(0)}_{u,b,k}$ we use the notation
\begin{flalign}\label{SINR_single_receiver}
\overline{\text{SINR}}_{u,b,k}(\boldsymbol P) = \frac{|h_{u,b,k}|^2\sum_{b\in\mathcal{B}_v}P_{u,b,k}}{\sigma^2_{b,k}+\sum_{\substack{\tilde{u}\in\mathcal{U}_v,\tilde{u}\neq u,\\\tilde{b}\in\mathcal{B}_v}} |h_{\tilde{u},b,k}|^2P_{\tilde{u},\tilde{b},k}}.
\end{flalign} 
The interference term in the denominator of (\ref{SINR_single_receiver}) incorporates the constraint that each user communicates with at most one BS at each frequency band. This constraint does not appear in the interference term of the SINR expression (\ref{eq:SINR_def}). This follows since the channel allocation is a by-product of the power allocation scheme presented in Section \ref{sec:joint_power_allocation}. That is, a user is allocated a channel only when the power allocation scheme allocates strictly positive power to  the transmission of the user over that channel.  

Next we present three channel allocation schemes. The first of these channel allocation schemes is a user-centric (UC) one  in which, at each frequency band, every user chooses its receiving BS to be the one  with the maximal SINR for this user given an initial power allocation. The second and third channel allocation schemes are existing approaches that we also consider for comparison. In particular, the second scheme is BS-centric (BSC) used, for example, in \cite{6678362,6815733}. In this scheme, in each frequency band every BS chooses its transmitting user to be the one with the maximal SINR. The third and final channel allocation scheme we consider
 is presented in \cite{7873307}. In this scheme, given a power allocation, channels are allocated to maximize the sum rate for that given power allocation using the Hungarian methods. We refer to this approach as the maximum sum rate matching (MSRM) approach.                
Interestingly,  numerical results show that, as the number of virtual cells decreases and their size increases, both the UC channel allocation and the equivalent continuous problem approach outperform both the BSC approach and the MSRM approach.  We remark that this work only considers a single power allocation scheme in Algorithm \ref{algo:Altenating_general}. That is due to the results presented in \cite{6678362}, where different power allocation schemes coupled with channel allocation yielded virtually the same average throughput. Hence we believe that different power allocation schemes will yield little difference in the system sum rate from that obtained with the power allocation algorithm used in Algorithm \ref{algo:Altenating_general}.

\subsection{User-Centric (UC) Channel Allocation}\label{sec:alternating_power_allocation_user}
This section presents the first  channel allocation scheme, depicted in Algorithm \ref{algo:Altenating_single_set_power_UCB}, for the interference coordination model. This scheme is a UC one in that every user chooses the receiving BS to be the one  with the maximal SINR for this user.

\setlength{\textfloatsep}{.7cm}
\begin{algorithm}
	\caption{}\label{algo:Altenating_single_set_power_UCB}
	\begin{algorithmic}[1]
		\State Input: Power allocation $\boldsymbol P = (P_{u,b,k})_{u\in\mathcal{U}_v,b\in\mathcal{B}_v,k\in\mathcal{K}}$;
		\State For every $u\in\mathcal{U}_v$, $b\in\mathcal{B}_v$ and $k\in\mathcal{K}$ calculate $\overline{\text{SINR}}_{u,b,k}(\boldsymbol P)$; 		
		\State For every $u\in\mathcal{U}_v$ and $k\in\mathcal{K}$, calculate: $b_{u,k} = \arg\max_{b\in\mathcal{B}_v} \overline{\text{SINR}}_{u,b,k}(\boldsymbol P)$;
		\State For every $(u,b,k)\in\mathcal{U}_v\times\mathcal{B}_v\times\mathcal{K}$ set $\gamma_{u,b,k}=\mathbbm{1}_{\{b = b_{u,k}\}}$;
	\end{algorithmic}
\end{algorithm}

The motivation behind this approach is allowing the power allocation stage more flexibility to choose the users who transmit to a given BS. More specifically,
in previously proposed channel allocation schemes discussed in Sections \ref{sec:alternating_power_allocation_BS} and \ref{sec:MSRM_channel_allocation}, at most one user is allocated to a BS in each frequency band. However, in the UC approach, in each frequency band each BS has a list of users that chose it as their receiving BS, then the power allocation stage chooses the identity of the user in that list who actually transmits to the BS by allocating to that user a positive transmission power.
Interestingly, numerical results show that as the number of virtual cells decreases and their size increases, both the UC channel allocation and the equivalent continuous problem approach outperform both of the previously-proposed channel allocation  methods that we next discuss.

\subsection{Base Station (BS) Centric Resource Allocation}\label{sec:alternating_power_allocation_BS}
This section presents the second channel allocation scheme for the interference coordination model. This scheme is a BS-centric one in that every BS chooses its transmitting user to be the one with the maximal SINR for this BS. This scheme is inspired by the works \cite{6678362} and \cite{6815733}, however, we remark that we do not restrict users in this work to transmit to the same BS over all frequency bands but allow them to communicate with different BSs in the virtual cell across different frequency bands. 
We depict the BS-centric channel allocation scheme in  Algorithm \ref{algo:Altenating_single_set_power_BCU}. 
\setlength{\textfloatsep}{.7cm}
\begin{algorithm}
	\caption{}\label{algo:Altenating_single_set_power_BCU}
	\begin{algorithmic}[1]
		\State Input: Power allocation $\boldsymbol P = (P_{u,b,k})_{u\in\mathcal{U}_v,b\in\mathcal{B}_v,k\in\mathcal{K}}$;
		\State For every $u\in\mathcal{U}_v$, $b\in\mathcal{B}_v$ and $k\in\mathcal{K}$ calculate $\overline{\text{SINR}}_{u,b,k}(\boldsymbol P)$;		
		\State For every $b\in\mathcal{B}_v$ and $k\in\mathcal{K}$, calculate: $u_{b,k} = \arg\max_{u\in\mathcal{U}_v} \text{SINR}_{u,b,k}(\boldsymbol P)$;
		\State For every $u\in\mathcal{U}_v$, $b\in\mathcal{B}_v$ and $k\in\mathcal{K}$ set $\gamma_{u,b,k}=\mathbbm{1}_{\{u = u_{b,k}\}}$;
	\end{algorithmic}
\end{algorithm}

The motivation behind this approach is interference reduction, that is, if the SINR at two or more BSs is maximized by the same user, then a transmission of this user intended for one of these BSs strongly interferes with the communication of the other BS. To reduce interference, the same user is chosen as the transmitting user by all of these BSs, then the power allocation scheme will chose the identity of the receiving BSs among them in accordance with the global objective function of the power allocation stage.    

We remark that even though in this approach several BSs can choose the same user, it can be proved, following the argument presented in the proof of Theorem \ref{theorem:equivalence:discrete_continuous}, that an optimal power allocation scheme  will allocate power only to the transmission of no more than one BS. In practice, this behavior is  observed using the high SINR approximation.
If  a power allocation scheme that does not display this behavior is used, that is, after the power allocation stage there is a user that has a positive transmission power over the same frequency band to two or more BSs, one can improve the  sum rate by using all the allocated transmit power of that user over that frequency band to the communication with the BS that has the highest SINR for that frequency band.

\subsection{Maximum Sum Rate Matching (MSRM) Channel Allocation}\label{sec:MSRM_channel_allocation} 

This section presents the third and final channel allocation scheme for the interference coordination model. This scheme allocates the channels in a virtual cell optimally for a given power allocation by solving the maximum sum rate matching problem; this approach is presented in \cite{7873307}.  
Next we depict the channel allocation problem as a matching problem.  
Let $B_k=(\mathcal{U}_v,\mathcal{B}_v,E,\boldsymbol{P},k)$ denote the bipartite graph that connects the set of users $\mathcal{U}_v$ to the set of BSs $\mathcal{B}_v$ where the set $E$ is the set of all pairs $\{u,b\}$ such that $u\in\mathcal{U}_v$ and $b\in\mathcal{B}_v$. Each edge $\{u,b\}$ is assigned a weight that is equal to the transmission rate from $u$ to $v$ using frequency band $k$, given the power allocation $\boldsymbol{P}$. We  allocate the channels at each frequency band $k$ by solving the sum rate maximization matching problem of $B_k$ optimally.  This optimal matching  can be found for example by using the Hungarian method \cite{doi:10.1002/nav.3800020109} for every $B_k$.  
This channel allocation scheme is depicted in Algorithm \ref{algo:Altenating_single_set_power_assignment}.
\setlength{\textfloatsep}{.7cm}
\begin{algorithm}
	\caption{}\label{algo:Altenating_single_set_power_assignment}
	\begin{algorithmic}[1]
		\State Input: Power allocation $\boldsymbol P = (P_{u,b,k})_{u\in\mathcal{U}_v,b\in\mathcal{B}_v,k\in\mathcal{K}}$;
		\State For every $u\in\mathcal{U}_v$, $b\in\mathcal{B}_v$ and $k\in\mathcal{K}$ calculate $\overline{\text{SINR}}_{u,b,k}(\boldsymbol P)$
		and
		\[R_{u,b,k}=W_k\log_2\left(1+\text{SINR}_{u,b,k}(\boldsymbol P)\right);\] 
		\State For every $k\in\mathcal{K}$ find the optimal matching of $B_k=(\mathcal{U}_v,\mathcal{B}_v,E,\boldsymbol{P},k)$, then
		set $\gamma_{u,b,k}=1$ if user $u$ was matched with BS $b$ in frequency band $k$ and $\gamma_{u,b,k}=0$ otherwise;
	\end{algorithmic}
\end{algorithm}

We note that, as stated in \cite{7873307}, given a power allocation $\boldsymbol P$,  Algorithm \ref{algo:Altenating_single_set_power_assignment} finds the optimal channel allocation that maximizes the sum rate for that power allocation. However, since the power allocation may not be optimal, the overall solution is not necessarily optimal. Interestingly, as we previously wrote, numerical results show that as the number of virtual cells decreases and their size increases both the user-centric channel allocation and the equivalent continuous problem approach outperforms this scheme.     

\subsection{Convergence of Algorithm \ref{algo:Altenating_general}}

The convergence of Algorithm \ref{algo:Altenating_general}  depends on the channel allocation scheme used and the initial values  $\alpha_{u,b,k}^{(0)}$. Since the system sum rate is bounded, convergence must occur whenever there is an $N_0\in\mathbb{N}$ such that $R(\boldsymbol P^{(n)},\boldsymbol\gamma^{(n)})\geq R(\boldsymbol P^{(n-1)},\boldsymbol\gamma^{(n-1)})$ for all $n\geq N_0$. This, in turn must occur if 
$R(\boldsymbol P^{(n-1)},\boldsymbol\gamma^{(n)})\geq R(\boldsymbol P^{(n-1)},\boldsymbol\gamma^{(n-1)})$ and $R(\boldsymbol P^{(n)},\boldsymbol\gamma^{(n)})\geq R(\boldsymbol P^{(n-1)},\boldsymbol\gamma^{(n)})$ for every $n\geq N_0$. This condition holds  when allocating channels using Algorithm \ref{algo:Altenating_single_set_power_UCB} or Algorithm  \ref{algo:Altenating_single_set_power_assignment} and choosing the initial values $\alpha_{u,b,k}^{(0)}$ at time $n$ to be $\gamma^{(n)}_{u,b,k}\overline{\text{SINR}}_{u,b,k}(\boldsymbol P^{(n-1)})$, since Algorithm \ref{algo:Altenating_single_set_power_UCB} and Algorithm \ref{algo:Altenating_single_set_power_assignment} cannot decrease the sum rate of a virtual cell, and since the high SINR approximation (\ref{eq:high_SINR_approx_improved}) is achieved with equality for $z=z_0$. In practice, convergence was observed in simulations for all channel allocation algorithms presented in this work for the choices $\alpha_{u,b,k}^{(0)}=\gamma^{(n)}_{u,b,k}\overline{\text{SINR}}_{u,b,k}(\boldsymbol P^{(n-1)})$ and $\alpha_{u,b,k}^{(0)}=\gamma^{(n)}_{u,b,k}$. The latter
 choice provided a small improvement over the first and was used in our simulations. 

\section{Resource Allocation for Coordinated Multi-Point Decoding in Virtual Cells}\label{sec:joint_decoding}
This section is dedicated to solving the problem (\ref{eq:uplink_problem_clean}) that is presented in Section \ref{subsection:uplink_joint_decoding_problem}
in which BSs use cloud decoding with backhaul links of infinite capacity.  Note that this  setup is equivalent to a multiple access channel (MAC) with a single transmitting antenna at each user and  multiple antennas at the receiver. 

Using the identity $\det(\boldsymbol{AB})=\det(\boldsymbol{A})\det(\boldsymbol{B})$ we have that  problem (\ref{eq:uplink_problem_clean}), which depicts the capacity of the virtual cell, can be written as follows: 
\begin{flalign}\label{problem_joint_decode_ininite}
\max &\sum_{k\in\mathcal{K}}W_k\left[\log_2\det\left(\boldsymbol{N}_{v,k}+\sum_{u\in\mathcal{U}_v}p_{u,k}\boldsymbol h_{u,k} \boldsymbol h_{u,k}^{\dagger}\right)-\log_2\det\left(\boldsymbol{N}_{v,k}\right)\right],\nonumber\\
\text{s.t.: } & \sum_{k\in\mathcal{K}} p_{u,k}\leq \overline{P}_{u,k},\quad p_{u,k}\geq 0.
\end{flalign}
Since the terms $\log_2\det\left(\boldsymbol{N}_{v,k}\right)$ are constants, hereafter we omit them from the objective function.

Denote $\boldsymbol{p}_u = (p_{u,1},\ldots,p_{u,K})$ and let:
\begin{flalign}
&f\left(\boldsymbol{p}_{u_1},\ldots,\boldsymbol{p}_{u_{|U|}}\right) =\log_2\det\left(\boldsymbol{N}_{v,k}+\sum_{u\in\mathcal{U}_v}p_{u,k}\boldsymbol h_{u,k} \boldsymbol h_{u,k}^{\dagger}\right).
\end{flalign} 

In order to optimally solve the problem (\ref{problem_joint_decode_ininite}) iteratively using the cyclic  coordinate ascend algorithm \cite[Chapter 2.7]{Bertsekas/99}, the following three conditions must hold:
\begin{enumerate}
	\item The function $f\left(\boldsymbol{p}_{u_1},\ldots,\boldsymbol{p}_{u_{|\mathcal{U}_v|}}\right)$ is concave.
	\item Define 
	\begin{flalign}
	\mathcal{P}&\triangleq  \left\{\left(\boldsymbol{p}_{u_1},\ldots,\boldsymbol{p}_{u_{|\mathcal{U}_v|}}\right):\sum_{k\in\mathcal{K}} p_{u,k}\leq \overline{P}_u,\:\: \sum_{k\in\mathcal{K}}p_{u,k}\geq 0 \:\: \forall\: u\in\mathcal{U}_v\right\},\nonumber\\
	\mathcal{P}_u&\triangleq\left\{\boldsymbol{p}_u:\sum_{k\in\mathcal{K}}p_{u,k}\leq \overline{P}_u,\:p_{u,k}\geq0\right\},
	\end{flalign} 
	then $\mathcal{P} = \mathcal{P}_{u_1}\times\ldots\times\mathcal{P}_{u_{|U|}}$.
	\item The problem
	\begin{flalign}\label{problem_joint_decode_ininite_single}
	\max_{\tilde{\boldsymbol{p}}_{u_i}}\: &f\left(\boldsymbol{p}_{u_1},\ldots,\boldsymbol{p}_{u_{i-1}},\tilde{\boldsymbol{p}}_{u_i},\boldsymbol{p}_{u_{i+1}},\boldsymbol{p}_{u_{|U|}}\right)\nonumber\\
	\text{s.t.: } & \tilde{\boldsymbol{p}}_{u_i}\in\mathcal{P}_{u_i},
	\end{flalign}
	has a unique maximizing solution.
\end{enumerate}

Next we solve problem (\ref{problem_joint_decode_ininite_single}) and show that the  optimal solution is uniquely attained.

Denote $\boldsymbol\Sigma_{i,k} = \boldsymbol{N}_{v,k}+\sum_{j\neq i}p_{u_j,k}\boldsymbol h_{u_j,k}\boldsymbol h_{u_j,k}^{\dagger}$. Problem  (\ref{problem_joint_decode_ininite_single})
is then
\begin{flalign}\label{problem_joint_decode_ininite_single_eq}
\max &\sum_{k\in\mathcal{K}}W_k\log_2\det\left(\boldsymbol\Sigma_{i,k}+p_{u_i,k}\boldsymbol h_{u_i,k} \boldsymbol h_{u_i,k}^{\dagger}\right)\nonumber\\
\text{s.t.: } & \sum_{k\in\mathcal{K}} p_{u_i,k}\leq \overline{P}_{u_i},\quad p_{u_i,k}\geq 0.
\end{flalign}

The Lagrangian of (\ref{problem_joint_decode_ininite_single_eq}) is:
\begin{flalign*}
&L(\boldsymbol p_{u_i},\lambda,\boldsymbol\mu) =  \sum_{k\in\mathcal{K}}W_k\log_2\det\left(\boldsymbol\Sigma_i(k)+p_{u_i,k}\boldsymbol h_{u_i,k} \boldsymbol h_{u_i,k}^{\dagger}\right)-\lambda_{u_i}(\sum_{k\in\mathcal{K}} p_{u_i,k}-\overline{P}_{u_i})
+\sum_{k\in\mathcal{K}}\mu_{u_i,k}p_{u_i,k}.
\end{flalign*}

Next, we calculate the derivative of the Lagrangian with respect to $p_{u_i,k}$:
\begin{flalign}
\frac{\partial L(\boldsymbol p_{u_i},\lambda,\boldsymbol\mu)}{\partial p_{u_i,k}}&=  W_k\boldsymbol h_{u_i,k}^{\dagger}\left(\boldsymbol\Sigma_{i,k}+p_{u_i,k}\boldsymbol h_{u_i,k} \boldsymbol h_{u_i,k}^{\dagger}\right)^{-1}\boldsymbol h_{u_i,k} - \lambda_{u_i}+\mu_{u_i,k} \nonumber\\
& = W_k\frac{\boldsymbol h_{u_i,k}^{\dagger}\boldsymbol\Sigma_{i,k}^{-1}\boldsymbol h_{u_i,k}}{1+\boldsymbol h_{u_i,k}^{\dagger}\boldsymbol\Sigma_{i,k}^{-1}\boldsymbol h_{u_i,k}p_{u_i,k}}-\lambda_{u_i}+\mu_{u_i,k}.
\end{flalign} 

The KKT conditions for (\ref{problem_joint_decode_ininite_single_eq}) are
\begin{flalign}
& W_k\frac{\boldsymbol h_{u_i,k}^{\dagger}\boldsymbol\Sigma_{i,k}^{-1}\boldsymbol h_{u_i,k}}{1+\boldsymbol h_{u_i,k}^{\dagger}\boldsymbol\Sigma_{i,k}^{-1}\boldsymbol h_{u_i,k}p_{u_i,k}}-\lambda_{u_i}+\mu_{u_i,k} = 0,\nonumber\\
& \lambda_{u_i}\left(\sum_{k\in\mathcal{K}} p_{u_i,k}-\overline{P}_{u_i}\right) = 0,\quad \mu_{u_i,k}p_{u_i,k}=0,\nonumber\\
& \mu_{u_i,k}\geq 0,\quad \lambda_{u_i}\geq0.
\end{flalign}

Since $\mu_{u_i,k}$ is nonnegative for all $k$, and the matrix $\boldsymbol\Sigma^{-1}_{i,k}$ is positive definite for all $k$, in order to fulfill 
the first KKT condition, $\lambda_{u_i}$ must be strictly positive.  

Now, if $p_{u_i,k}>0$, then $\mu_{u_i,k} =0 $ and by the first KKT condition we have
\begin{flalign}
p_{u_i,k}=\frac{W_k}{\lambda_{u_i}}-\frac{1}{\boldsymbol h_{u_i,k}^{\dagger}\boldsymbol\Sigma_{i,k}^{-1}\boldsymbol h_{u_i,k}}.
\end{flalign} 
Also, if $p_{u_i,k}=0$, then by the first KKT condition we have
\begin{flalign}
W_kh_{u_i,k}^{\dagger}\boldsymbol\Sigma_{i,k}^{-1}\boldsymbol h_{u_i,k}+\mu_{u_i,k} = \lambda_{u_i}.
\end{flalign} 

It follows that 
\begin{flalign}
p_{u_i,k} =  \left(\frac{W_k}{\lambda_{u_i}}-\frac{1}{\boldsymbol h_{u_i,k}^{\dagger}\boldsymbol\Sigma_{i,k}^{-1}\boldsymbol h_{u_i,k}}\right)^+
\end{flalign}
where $\lambda_{u_i}$ is chosen such that $\sum_{k\in\mathcal{K}}p_{u_i,k} = \overline{P}_{u_i}$.

\section{Numerical Results}\label{se:simulation}
This section presents Monte Carlo simulation results for the resource allocation and user affiliation schemes  presented in this paper.  In these simulations there are $8$ frequency bands, each of bandwidth 20 KHz, and the carrier frequency is set to $1800$ MHz. The noise power received by each BS is $-174$ dBm/Hz, and the maximal power constraint for each user is $23$ dBm. Finally, in each frequency band the channel exhibits Rayleigh fading, log-normal shadowing with standard deviation $8$ dB, and a path loss of $PL(d)= 35\log_{10}(d)+34$, where $d$ denotes the distance between the transmitter and the receiver in meters (see \cite{4138008}). 
The network comprises $15$ BSs and $100$ users which are uniformly located in a square of side $2000$ meters. The results are averaged over $1000$ system realizations.
The numerical results depict the average system sum rate achieved by the BS clustering,  resource allocation methods, and user affiliation scheme we propose in this paper. To evaluate the performance of our BS clustering we compare the average system-sum rate achieved by using the hierarchical clustering with minimax linkage criterion to that of other popular clustering algorithms, namely, the K-means clustering algorithm and the spectral clustering  algorithm \cite{Ng:2001:SCA:2980539.2980649} for the choices $\sigma=\sqrt{2000}$ and $\sigma=2000$. 
The simulation results for the system setup stated above are shown in Figures \ref{Best_Channel_Average_Sum_Rate_fig_single_all}-\ref{Several_Joint_decoding_exhaustive_hierarchial_fig}. An additional figure, Fig.~\ref{Several_Comparison_Clustering_Max_Average_Sum_Rate_max_single2}, presents numerical results that evaluate the clustering choice for a system setup with $10$ BSs and $80$ users that are uniformly located in a square of side $1000$ meters; all the other system parameter remain the same.

The line descriptions of the figures are of the structure $F1 - F2 - F3$ where
\begin{itemize}
\item The $F1$  field  describes the BS clustering method used. This field can take one of following options: \textit{Hierarchical}, which stands for the hierarchical clustering with minimax linkage criterion; \textit{K-means}, which stands for the K-means clustering algorithm, and \textit{Spectral clustering $\sigma=x$}, which stands for spectral clustering where $\sigma$ takes the value $x$.
\item The $F2$ field describes the resource allocation scheme. This field can take one of the following options:
\begin{itemize}
\item \textit{JD}, which stands for Joint Decoding, refers to the resource allocation schemes for the coordinated multi-point model which is presented in Section \ref{sec:joint_decoding}. 
	\item \textit{Continuous}, which refers to the resource allocation  presented in Section \ref{sec:joint_power_allocation}.
	\item  \textit{UC}, which refers to the resource allocation presented in Section  \ref{sec:alternating_power_allocation_user}.
	\item \textit{BSC}, which  refers to the resource allocation presented in Section \ref{sec:alternating_power_allocation_BS}.
	\item \textit{MSRM}, which  refers to the resource allocation presented in Section \ref{sec:MSRM_channel_allocation}.
	\item \textit{Max SUD} which refers to the maximal average sum rate produced by each of the above resource allocation schemes for the interference coordination model.
	\end{itemize}
\item The $F3$ field describes the user affiliation criterion. This field
 can either be \textit{``best channel"} or \textit{``closest BS"}.
\end{itemize}

\subsection{Average System Sum Rate}

\begin{figure}
 	\centering
 	\includegraphics[scale=0.67]{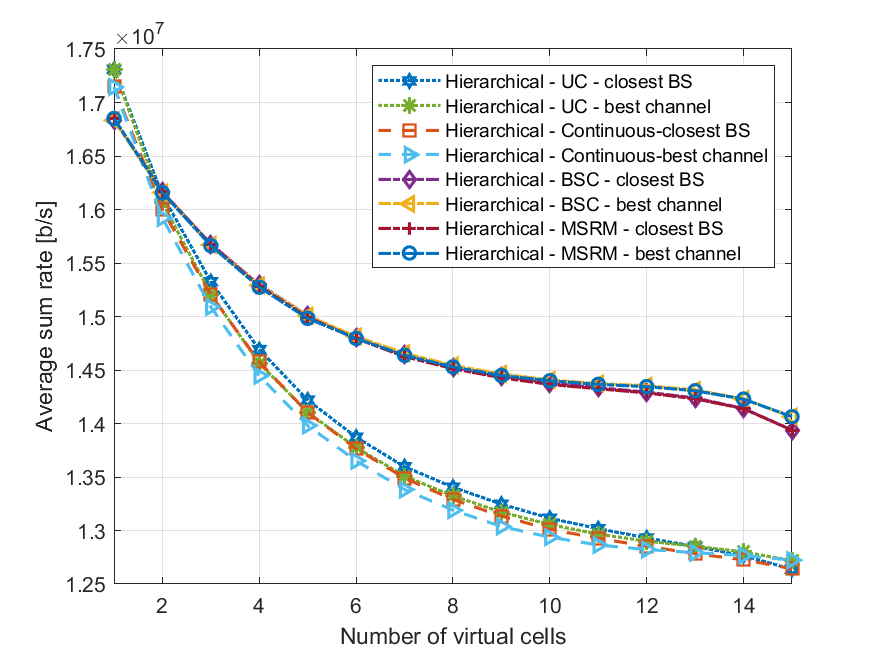}
 	\vspace{-0.4cm}
 	\caption{Comparison of the average system sum rate of the interference coordination model  as a function of the number of virtual using hierarchical BS clustering with minimax linkage criterion.}
 	\label{Best_Channel_Average_Sum_Rate_fig_single_all}
 	\vspace{-0.3cm}
 \end{figure}

\begin{figure}
 	\centering
 	\includegraphics[scale=0.67]{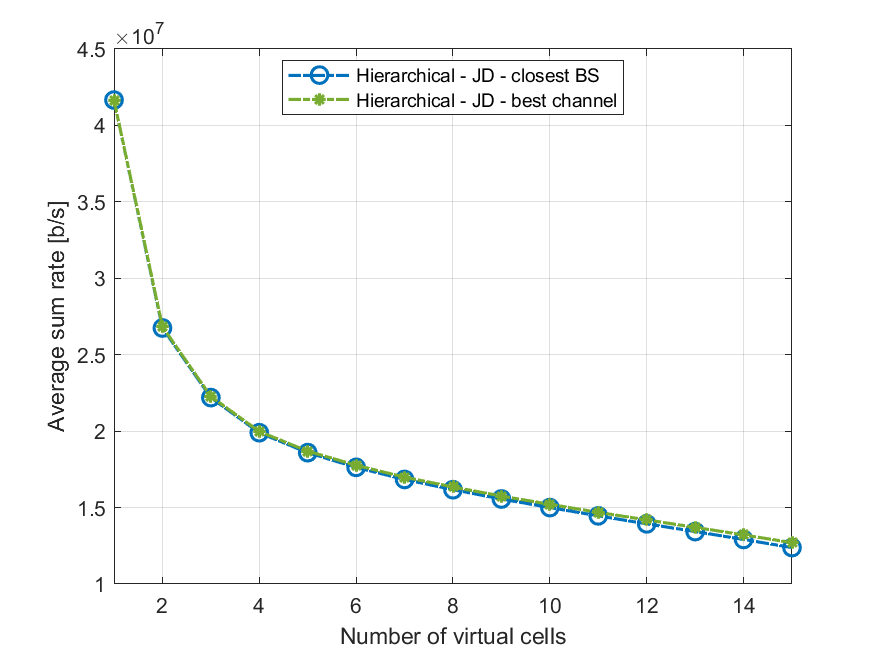}
 	\vspace{-0.4cm}
 	\caption{Comparison of the average system sum rate of the coordinated multi-point decoding as a function of the number of virtual cells using hierarchical BS clustering with minimax linkage criterion.}
 	\label{Best_Channel_Average_Sum_Rate_joint_decoding}
 	\vspace{-0.3cm}
 \end{figure}

\begin{figure}	
 	\centering
 	\includegraphics[scale=0.67]{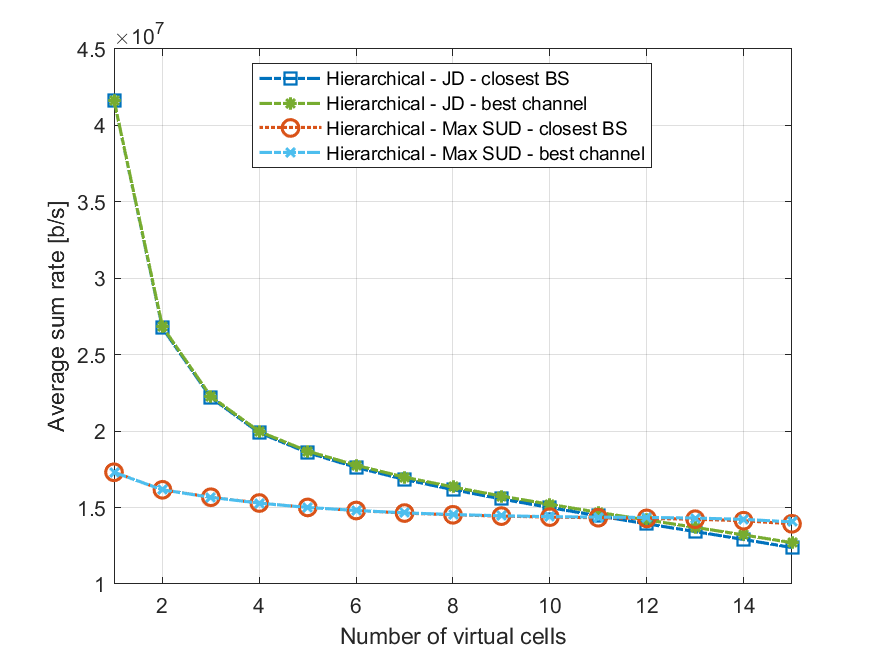}
 	\vspace{-0.4cm}
 	\caption{Comparison between the average sum rate of the interference coordination model and the coordinated multi-point decoding  as a function of the number of virtual cells using hierarchical BS clustering with minimax linkage criterion.}
 	\label{Best_Channel_Average_Sum_Rate_fig_all}
 	\vspace{-0.4cm}
 \end{figure}

Figures \ref{Best_Channel_Average_Sum_Rate_fig_single_all}-\ref{Best_Channel_Average_Sum_Rate_fig_all} depict the average system sum rate as a function of the number of virtual cells.
We clustered the BSs in the network according to the hierarchical clustering algorithm with the minimax linkage criterion that is depicted in Algorithm \ref{algo:hierarchical_clustering}. We considered both of the user affiliation rules we propose in Section \ref{sec_user_affil}, i.e., the ``closest BS" criterion and the ``best channel" criterion.  
We examined the average system sum rate of both cooperation models discussed in this paper: the interference coordination model whose resource allocation schemes are discussed in Sections \ref{sec:joint_power_allocation}-\ref{sec:alternating_optimization}, and the coordinated multi-point decoding model whose resource allocation scheme is discussed in  Section \ref{sec:joint_decoding}.    
Fig.~\ref{Best_Channel_Average_Sum_Rate_fig_single_all} depicts the average system sum rate of the interference coordination model  for each of the resource allocation schemes and each of the user affiliation schemes we propose in this paper. 
 Fig.~\ref{Best_Channel_Average_Sum_Rate_joint_decoding}  depicts the average system sum rate of the coordinated multi-point decoding for  each of the user affiliation schemes we propose. Finally, Fig.~\ref{Best_Channel_Average_Sum_Rate_fig_all} depicts the average system sum rate achieved by  each of the cooperation models we consider.

  Figures \ref{Best_Channel_Average_Sum_Rate_fig_single_all}-\ref{Best_Channel_Average_Sum_Rate_fig_all}  lead to several interesting insights and conclusions. 
  First, they confirm the expectation that, as the number of virtual cells decreases, the average sum rate increases.  
  Second, they show that the best channel affiliation rule outperforms the closest BS one when the number of virtual cells is large. However, as Fig.~\ref{Best_Channel_Average_Sum_Rate_fig_single_all} shows, this changes in the interference coordination model when the number of virtual cells decreases. In this case the closest BS affiliation rule either outperforms or is on par with the best channel one, depending on the resource allocation scheme. 
  
  Additionally, Fig.~\ref{Best_Channel_Average_Sum_Rate_fig_single_all} shows that it is best to use the BSC or MSRM channel allocation methods, which yielded similar performance,  for allocating channels and power in virtual cells except when there is a single virtual cell (fully centralized optimization). In this case the two new resource allocation techniques that we propose outperform these other methods.
  This can be explained by the fact that our new schemes  provide more freedom in the power allocation stage to choose which users have a positive transmission power compared with existing methods. However, since the power allocation problem is solved approximately, its solution may not be optimal. When the size of the virtual cells is small (i.e. there are many virtual cells), the channel allocation choice of the existing methods is good whereas the new methods suffer loss in performance due to the suboptimality of the power allocation stage. However, as the size  of the virtual cells grows (as their number is decreased), the ability of the new methods to consider in the power allocation stage more channel allocation combinations improves the resource allocation performance, even though the solution of the power allocation problem is only approximately optimal.
   Overall the average sum rate increase of the resource allocation schemes of the fully centralized scenario, i.e., a single virtual cell compared to the fully distributed scenario, is approximately $20\%$ when considering the best achieved average sum rate at each point.

 Fig.~\ref{Best_Channel_Average_Sum_Rate_joint_decoding} depicts the average system sum rate of the coordinated multi-point decoding as a function of the number of virtual cells comprising the network. It shows the monotonic and significant improvement in average system sum rate as the number of virtual cells decreases; the overall improvement in average system sum rate is  330\%.

Fig.~\ref{Best_Channel_Average_Sum_Rate_fig_all} compares the average system sum rate achieved by the coordinated multi-point decoding and the one achieved by the interference coordination model. Fig.~\ref{Best_Channel_Average_Sum_Rate_fig_all}  shows that coordinated multi-point decoding can achieve significantly higher average system sum rate compared with single user decoding. However, single user decoding may yield a higher sum rate when the number of virtual cells is large. For a large number of virtual cells, where the limited coordination between BSs is similar to having no coordination between BSs, ignoring out of cell interference affects the joint decoding scheme more severely, since it depends on the exact second order statistics of the interference. Thus ignoring the interference outside virtual cells affects the coordinated multi-point scheme more severely than the interference coordination model with a large number of virtual cells. In this case  the loss in performance caused by using an inexact interference covariance matrix is not compensated by the gain in performance of using joint decoding in the virtual cell.

\subsection{Comparison with Other Clustering Algorithms}
\begin{figure}
	\centering
	\includegraphics[scale=0.67]{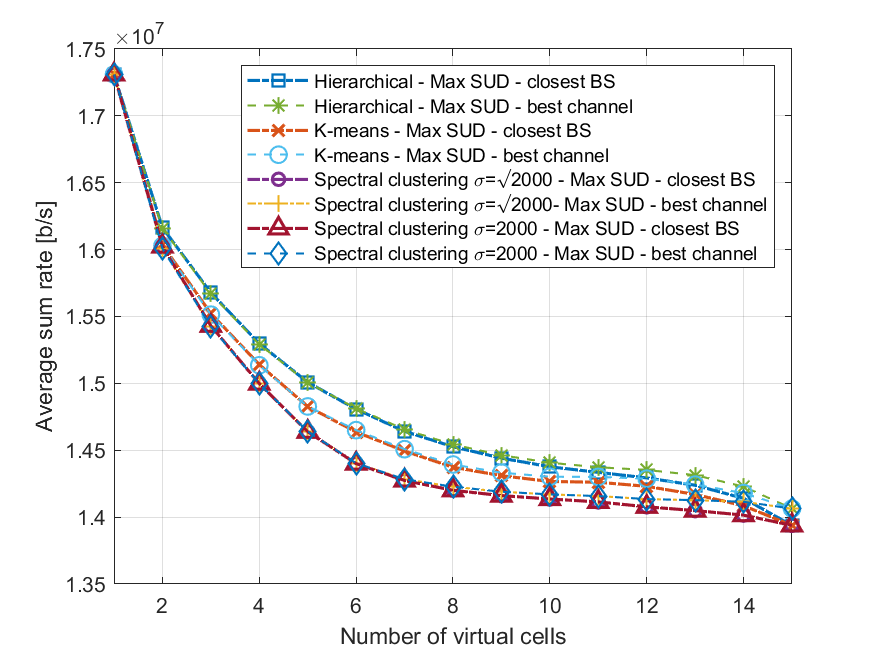}
					\vspace{-0.4cm}
	\caption{Comparison of the maximal average sum rate of several BSs clustering algorithms as a function of the number of virtual cells  for the interference coordination model.}
	\label{Several_Comparison_Clustering_Max_Average_Sum_Rate_max_single}
		\vspace{-0.4cm}
\end{figure}

\begin{figure}
	\centering
	\includegraphics[scale=0.67]{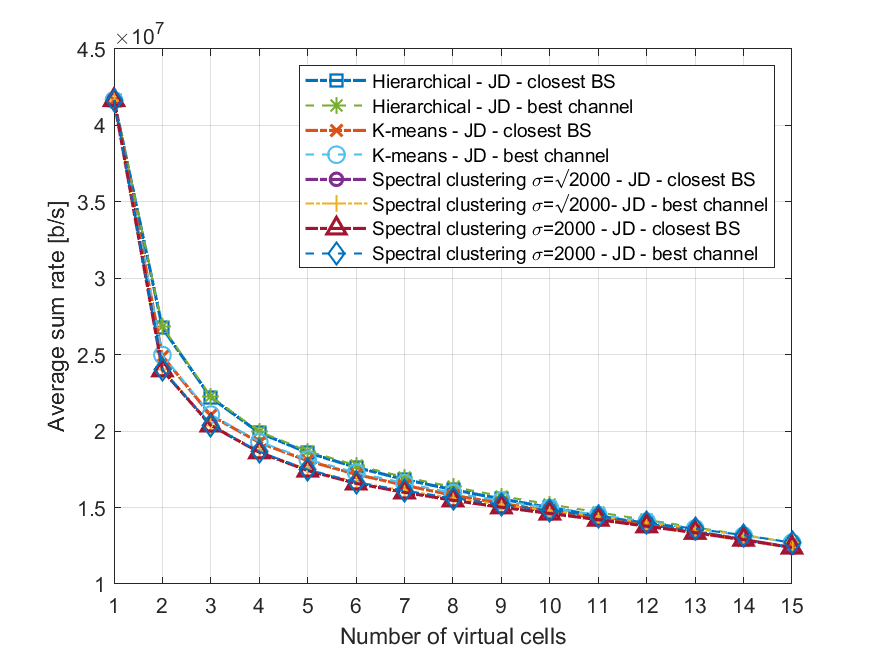}
					\vspace{-0.4cm}
	\caption{Comparison of the maximal average sum rate of several BSs clustering algorithms as a function of the number of virtual cells  for coordinated multi-point decoding.}
		\label{Several_Joint_decoding_exhaustive_hierarchial_fig}
			\vspace{-0.4cm}
\end{figure}

We also compared the average system sum rate using the hierarchical clustering algorithm with minimax linkage criterion with that of two other popular clustering algorithms, namely,  the K-means clustering algorithm and that of the spectral clustering  algorithm \cite{Ng:2001:SCA:2980539.2980649} for the choices $\sigma=\sqrt{2000}$ and $\sigma=2000$.  Fig.~\ref{Several_Comparison_Clustering_Max_Average_Sum_Rate_max_single} depicts the maximal average system sum rate achieved by each of the clustering algorithms where the maximization is taken over the resource allocation schemes for the interference coordination model presented in this work. Additionally, Fig.~\ref{Several_Joint_decoding_exhaustive_hierarchial_fig} depicts the  average system sum rate achieved by coordinated multi-point decoding.
Fig.~\ref{Several_Comparison_Clustering_Max_Average_Sum_Rate_max_single}-\ref{Several_Joint_decoding_exhaustive_hierarchial_fig} show that the hierarchical algorithm consistently outperforms both the K-means and the spectral clustering algorithms for both user affiliation rules and both cooperation models.

\begin{figure}
	\centering
	\includegraphics[scale=0.6]{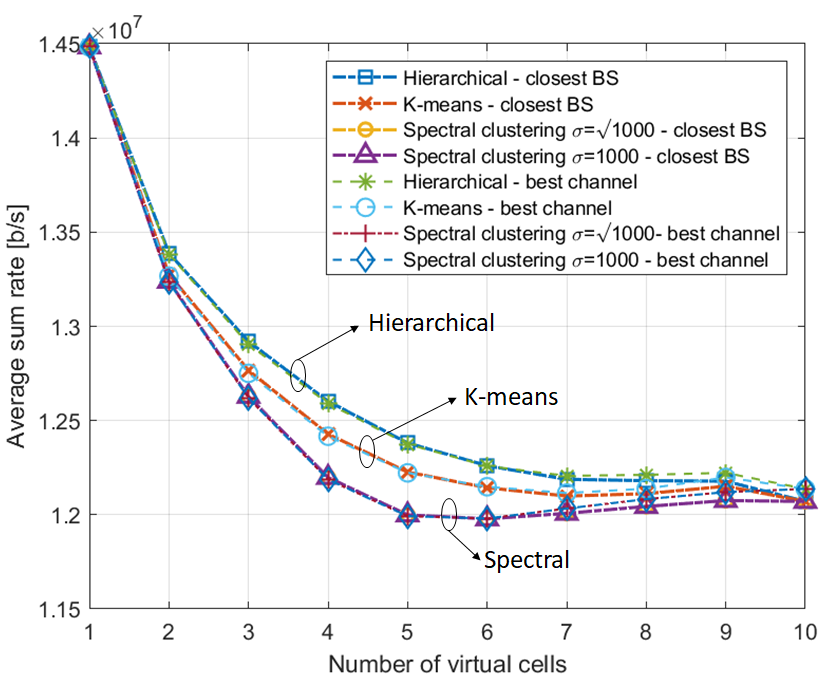}
					\vspace{-0.4cm}
	\caption{Comparison of the maximal average sum rate of several BSs clustering algorithms as a function of the number of virtual cells  for the interference coordination model.}
	\label{Several_Comparison_Clustering_Max_Average_Sum_Rate_max_single2}
		\vspace{-0.4cm}
\end{figure}

We considered an additional network setup which was comprised of $10$ BSs and $80$ users that were uniformly located in a square of side $1000$ meters. Fig.~\ref{Several_Comparison_Clustering_Max_Average_Sum_Rate_max_single2} presents the average system sum rate as a function of the number of virtual cells for the interference coordinated model. The results were averaged over $1000$ system realizations. Fig.~\ref{Several_Comparison_Clustering_Max_Average_Sum_Rate_max_single2} shows that a proper choice of  the clustering algorithm is crucial for improving network performance. This is evident in the plot of the spectral clustering algorithm in which the network performance monotonically decreases as the number of virtual cells is decreased from 10  to 5.

\section{Conclusion}\label{sec:conclusion}
This work addressed the role of virtual cells in resource allocation and network management for future wireless networks. It proposed methods for two design aspects of this network optimization; namely, forming the virtual cells and allocating the communication resources in each virtual cell to maximize total system sum rate. We considered two cooperation models in virtual cells. The first model used interference coordination, where the resource allocation in each virtual cell is performed jointly for all users and BSs in the virtual cell but there is no decoding cooperation. The second cooperation model  we considered was the coordinated multi-point decoding model, whereby BSs in a virtual cell allocate the communication resources jointly and also decode their signal cooperatively.  We presented two types of resource allocation schemes for the interference coordination model. The first scheme converted the NP-hard mixed-integer resource allocation problem into a continuous resource allocation problem and then found an approximate solution. The second scheme alternated between the power allocation and channel allocation problems. We proposed a new channel allocation that was carried out in a user-centric manner, and also considered a BS centric approach.  We additionally considered a maximum sum rate matching approach where an optimal channel assignment is found for a given power allocation. Since this power allocation may not be optimal, the overall solution may be sub-optimal as well.  We also solved the joint decoding resource allocation problem for the coordinated multi-point decoding model in each virtual cell optimally.  All of these schemes assume the BSs have been assigned to virtual cells via clustering. For this clustering we proposed the use of  hierarchical clustering  in the clustering of the BSs to form the virtual cells, since changing the number of virtual cells only causes local changes and does not force a reclustering of all the virtual BSs in the network. We presented numerical results for all of the aforementioned models. Our numerical results demonstrate the increase in system sum rate that our neighborhood-based optimization yields. This increase is monotonic as the neighborhood-based optimization reverts from distributed to centralized optimization.  Additionally, our numerical results indicate that coordinated multi-point communication systems show greater increase in system sum rate as the number of virtual cells decreases, in comparison with interference coordination communication systems.  Finally, they show that the hierarchical clustering with the minimax linkage criterion yields higher system sum rate than both K-means and spectral clustering. 
 
\bibliographystyle{IEEEtran}



\end{document}